\def\o{\over}
\def\A{\rightarrow}
\def\bar{\overline}

\def\a{\alpha}
\def\b{\beta}
\def\n{\nu}
\def\m{\mu}

\def\e{\epsilon}

\def\th{\theta}

\def\Im{{\rm Im}}

\def\bar{\overline}

\def\G{{\rm GeV}}
\def\M{{\rm MeV}}
\def\eV{{\rm eV}}
\documentstyle [epsfig,12pt]{article}
\begin{document}
\baselineskip=24.5pt
\setcounter{page}{1}
\thispagestyle{empty}
\topskip 0.5  cm
\begin{flushright}
\begin{tabular}{c c}
& {\normalsize   EHU-96-6 Revised}\\
& August 1996
\end{tabular}
\end{flushright}
\vspace{1 cm}
\centerline{\Large\bf  Is  CP Violation  Observable in Long Baseline}
\centerline{\Large\bf   Neutrino  Oscillation Experiments ?}
\vskip 1.5 cm
\centerline{{\bf Morimitsu TANIMOTO}
  \footnote{E-mail address: tanimoto@edserv.ed.ehime-u.ac.jp}}
\vskip 0.8 cm
 \centerline{ \it{Science Education Laboratory, Ehime University, 
 790 Matsuyama, JAPAN}}
\vskip 2 cm
\centerline{\bf ABSTRACT}\par
\vskip 0.5 cm
  We have studied $CP$ violation originated by the phase of the neutrino
 mixing matrix in the long baseline  neutrino oscillation experiments.
The direct measurements of  $CP$ violation is
  the difference of the transition probabilities between $CP$-conjugate
    channels.
	In those experiments, the $CP$ violating effect is not suppressed 
	 if the highest neutrino mass scale is 
	  taken to be $1\sim 5 \eV$, which is appropriate for the cosmological
  hot dark matter.
	 Assuming the hierarchy for  the neutrino masses,
  the upper bounds  of $CP$ violation have been caluculated 
	  for three cases, in which mixings
	 are constrained by the recent short baseline ones.
     The calculated upper bounds are  larger than  $10^{-2}$, 
	 which will be observable
	 in the long baseline accelerator experiments.
  The matter effect, which is not $CP$ invariant,
       has been also estimated in those experiments.
 \newpage
\topskip 0 cm
\section{Introduction}
  
   The origin of $CP$ violation is still an open problem in particle physics.
   In the quark sector, $CP$ violation  has been intensively studied
    in the KM standard model \cite{KM}.
 For the lepton sector,  $CP$ violation is also expected unless
	  the neutrinos are  massless.
    In particular, $CP$ violation in the neutrino flavor oscillations is 
	an  important phenomenon because it relates directly 
	to the $CP$ violating phase parameter in the mixing matrix
   for the massive neutrinos  \cite{CP}.
   Unfortunately, this $CP$ violating effect is suppressed in the short 
 baseline accelerator
   experiments if the neurinos have the hierarchical mass spectrum.
   However, the  suppression is avoidable 
   in the long baseline accelerator experiments, which are expected to operate
   in the near future \cite{long} \cite{KEKPS}. 
  So one has a chance to observe the $CP$ violating effect
    in those experiments.\par
	
    The recent indications of a deficit in the $\n_{\m}$ flux of 
   the atomospheric neutrinos \cite{Atm1}-\cite{Kam}
      has renewed interest in using accelerator neutrinos
    to perform the long baseline neutrino oscillation experiments.
	Many possibilities of  experiments have been discussed \cite{long}.
   The purpose of this paper is to present the numerical study of 
   $CP$ violation  in those accelerator experiments.\par
	  There are only two hierarchical mass difference scales 
 $\Delta m^2$  in the three-flavor mixing scheme 
 without introducing sterile neutrinos. 
 If  the highest neutrino mass scale is  taken to be $1\sim 5 \eV$,
  which is appropriate for the cosmological hot dark matter(HDM) \cite{hdm},
 the other mass scale is
  either the atomospheric neutrino mass scale  $\Delta m^2\simeq 10^{-2}\eV^2$
  \cite{Atm1}-\cite{Kam}
 or the solar neutrino one $\Delta m^2\simeq 10^{-5}\sim 10^{-6}\eV^2$
  \cite{solar}.
  Since the long baseline  experiments correspond to 
  the atomospheric neutrino mass scale, 
we take  $\Delta m^2\simeq 10^{-2}\eV^2$ as the lower mass scale.
 The solar neutrino problem is not discussed in this paper.
  The solar one  may be solved by introducing
 the sterile neutrino \cite{sterile}.
  \par
  Our study of  $CP$ violation  is presented in the framework of above pattern
  of the neutrino mass spectrum
  \footnote{
 $CP$ and $T$ violations have been studied in the case of
  $\Delta m_{31}^2 \sim 10^{-2} \eV^2$ \cite{Ara}.}.
 We also investigate the matter effect in the long baseline accelerator
 experiments
 since the background matter effect is not $CP$ invariant.
 If the matter effect is not negligibly small compared to
 the $CP$ violating effect in the vacuum, one should consider 
 how to extract the matter effect from the data.
 It is found that the matter effect strongly depends 
 on the hierarchical pattern of the neutrino masses and mixings.
\\
 \section{CP Violation in Neutrino Flavor Oscillations} 
   
    The amplitude of  $\n_\a\A \n_\b$ transition with
	 the neutrino  energy $E$
   after traversing the distance $L$ can be written as
  \begin{equation}        
 {\cal{A}}(\n_\a\A \n_\b) = e^{-i EL} \left\{\delta_{\a\b} 
   + \sum_{k=2}^3  
 U_{\a k} U_{\b k}^* \left [\exp{\left (-i {\Delta m^2_{k1} L \o 2 E}\right )}
	   -1 \right ] \right\} \ , 
		 \label{Pro}
\end{equation}
\noindent  where  $\Delta m^2_{ij}=m^2_i-m^2_j$ is defined, and  
$U_{\a i}$ denote the elements of the $3\times 3$ neutrino  flavor 
mixing matrix, in which $\a$ and $i$  refer to the flavor eigenstate and 
 the mass eigenstate, respectively. 
  The  amplitude  ${\cal A} (\bar\n_\a\A \bar\n_\b)$ is given by 
  replacing $U$ with $U^*$ in the right hand side in eq.(\ref{Pro}).
 The direct measurements of $CP$ violation 
 originated by the phase of the neutrino mixing matrix are 
    the differences of the transition probabilities between $CP$-conjugate
    channels \cite{CP}:  
  \begin{eqnarray}        
  \Delta P\equiv P(\bar \n_\m \A \bar \n_e) - P(\n_\m \A \n_e) &=&
    P(\n_\m \A \n_\tau) -  P(\bar \n_\m \A \bar \n_\tau)  \nonumber \\
 = P(\bar \n_e \A \bar \n_\tau)- P(\n_e \A \n_\tau)    
   &=&  4 J^{\n}_{CP} (\sin D_{12}+\sin D_{23}+\sin D_{31}) \ ,
   \label{CP}
   \end{eqnarray}
  \noindent
   where  
  \begin{equation}
 D_{ij}=  \Delta m^2_{ij}{L\o 2E} \ ,
 \label{D}
   \end{equation}
 and
 $J^{\n}_{CP}$ is defined for the  rephasing  invariant quantity of  
 $CP$ violation   in the neutrino mixing  matrix as well as the one 
in the quark sector \cite{J}.
 In terms of the standard parametrization of the  mixing matrix \cite{PDG},
	we have
	 \begin{equation}
    J_{CP}^{\n}= \Im (U_{\m 3}U^*_{\tau 3}U_{\m 2}^* U_{\tau 2})=
	 s_{12} s_{23} s_{13} c_{12} c_{23} c_{13}^2 \sin\phi \ ,
	 \label{JCP}
   \end{equation}
   \noindent where $\phi$ is the $CP$ violating phase.
   The oscillatory terms are periodic in $L/E$ and  
   $D_{12}+D_{23}+ D_{31}=0$ is satisfied.\par

   For  the neutrino masses,   we expect  the typical hierarchical relation
    $\Delta m^2_{31}\gg \Delta m^2_{32}$
 or $\Delta m^2_{31}\gg \Delta m^2_{21}$  in order to
 guarantee two different mass scales.
 The former relation(hierarchy I)  corresponds to 
 $m_3\simeq m_2\gg m_1$ 
 and the latter one(hierarchy II)  to $m_3\gg m_2\gg m_1({\rm or}\simeq m_1)$. 
 The highest neutrino mass scale is taken to be 
  $m_{3}=1\sim 5 \eV$, which is appropriate for the cosmological
 HDM\cite{hdm}.
Those mass  hierarchies have been discussed
 in the context of
 the solar neutrino problem, 
  atomospheric neutrino one and HDM(as well as LSND \cite{LSND})
 by several authors  \cite{sterile}\cite{pattern}.
 Although their results seem to support  hierarchy I,  both cases of
  hierarchies I and II are studied in this paper.

  Recently, there are significant short baseline accelerator experiments 
  \cite{LSND}-\cite{CHONOM}, in which
  the value of $L/E$  is fixed for each one.
 For example,  $L=30{\rm m}$ and $E=36\sim 60\M$ 
  are taken for the  $\bar \n_\m \A \bar\n_e$ experiment at LSND \cite{LSND}
  and $L=800{\rm m}$ and $E=30\G$ 
   for the  $\n_\m\A\n_\tau$ experiment at CHORUS and MOMAD \cite{CHONOM}.
   In these experiments, the value of $D_{31}$($\simeq -D_{12}$) is 
   $1\sim 10$ with  $|D_{23}|\ll 1$ for hierarchy I
    ($D_{31}\simeq -D_{23}$ is $1\sim 10$ and  with  $|D_{12}|\ll 1$
   for hierarchy II).
   Therefore the factor $(\sin D_{12}+\sin D_{23}+\sin D_{31})$ is suppressed
  because two largest terms almost cancel  due to   opposite signs.
 Another term is still small.
   Then the $CP$ violating effect in eq.(\ref{CP}) is 
   significantly reduced due to this suppression.
   So one has  no chance to observe  $CP$ violation for the present
  in the short baseline neutrino oscillation experiments. \par
   
   However, the situation is very different 
   in  the long baseline accelerator experiments.
    Let us consider  the case of   hierarchy I for the neutrino masses.
    The oscillatory terms $\sin D_{12}$ and $\sin D_{31}$  can be
	  replaced by the average value $0$
 since the magnitude of  $D_{31}$($\simeq -D_{12}$) is $10^3\sim 10^4$.
    Then $CP$ violation is dominated by the  $\sin D_{23}$ term,
   which is not  small because of $|D_{23}|\simeq 1$.
	 The same situation is kept  for hierarchy II.
	 Thus  the $CP$ violating quantity $\Delta P$  is not suppressed 
	  unless    $J^{\n}_{CP}$ is very small.
	\par
	  
\section{Non-Suppression of $CP$ Violation in Long Baseline Accelerator
   Experiments} 
    The long baseline accelerator experiments are planed to operate
   in the near future \cite{long}\cite{KEKPS}.
   The most likely possiblities are in KEK-SuperKamiokande(250Km),
  CERN-Gran Sasso(730Km) and  Fermilab-Soudan 2(730Km)
   experiments(MINOS).
  The sensitivity of the observable transition probabilitiy is
   expected to be  $10^{-2}$.
    The average energy of the $\n_\m$ beams are approximately $1\G$,
   $6\G$ and $10\G$ at KEK-PS$(12\G)$, CERN-SPS$(80\G)$ and  Fermilab
	 proton accelerator$(120\G)$, respectively.
    We can estimate $S_{CP}\equiv \sin D_{12}+\sin D_{23}+\sin D_{31}$
	 in those experiments.
	 The  neutrino energy $E$ dependences of $S_{CP}$ are shown 
	 by solid curves 
	  for fixed  $L=250$Km in fig.(a) and for $L=730$Km in fig.(b),
 where  $S_{CP}$ is averaged over  energy speread of $20\%$ at
 the reference energy.
 Here $\Delta m^2_{31}=2.25\eV^2$ and $\Delta m^2_{32}(\Delta m^2_{21})=
	    10^{-2}\eV^2$   are taken in  hierarchy I(II).
 Although our results depend on  the value of $\Delta m^2_{31}$,    
 these  change only $5\%$ for $\Delta m^2_{31}=1\sim 25\eV^2$.
  We also show the  oscillation function $\sin^2(\Delta m^2_{32} L/4 E)$,
 which is important for the absolute value of the transition probability
	  $P(\n_\a \A \n_\b)$ at those experiments,
	   by dashed curves in fig.1.\par

\begin{center}
\unitlength=0.7 cm
\begin{picture}(2.5,1.5)
\thicklines
\put(-2,0){\framebox(6,1){\bf Fig. 1(a) and (b)}}
\end{picture}
\end{center}
\par
As seen in fig.1(a), the absolute value of  $S_{CP}$  is almost maximum
 at $E \simeq 1.3\G$.  However, it  depends on
		the atomospheric neutrino mass scale  $\Delta m^2_{32}$.
  The  $\Delta m^2_{32}$ dependence of 	$S_{CP}$  is obtained
   by replacing the axis  $E$ with 
   $E\times \Delta m^2_{32}/10^{-2}\eV^2 $ in fig.1.
   Since the $CP$ violating effect changes considerably
    according to  values of $E$,
 the new project PS($50\G$) at KEK is very significant for observing $CP$ 
violation.
	 Both CERN-Gran Sasso and  MINOS
   experiments are also important because of  the non-suppression $S_{CP}$  
    as seen in fig.1(b).  
	 \vskip 0.5 cm

 \section{Constraints of Mixings from Present Reactor and Accelerator Data} 
  
    Analyses of the three flavor neutrino oscillation
  have been presented  recently\cite{Mina}-\cite{FLS}. 
 In particular, the quantitative results by
 Bilenky et al.\cite{Bil} and  Fogli, Lisi and  Scioscia \cite{FLS}
 are the useful guide for getting constraints of the neutrino mixings.
  The upper bound of $J_{CP}^{\n}$ is estimated by using these
  constraints.\par
  
 Let us begin with  discussing  constraints from  the   reactor and
  accelerator disappearance experiments. 
Since no indications in favor of neutrino oscillations were found in 
these experiments, we only  get the  allowed regions 
in $(U^2_{\a i}, \Delta m^2_{31})$ parameter space.
     Bugey reactor experiment \cite{Bugey} and  CDHS \cite{CDHS}
 and CCFR \cite{CCFR}
 accelerator experiments give  bounds for the neutrino
 mixing parameters at the fixed value of $\Delta m^2_{31}$.
 We follow the analyses given by Bilenky et al.\cite{Bil}.
\par
  Since the $CP$ violating effect can be neglected 
   in those short baseline experiments as discussed in section 3, 
   we  use the following  formula without $CP$ violation
     for the probability in the disappearance experiments:
   \begin{equation}        
 P(\n_\a\A \n_\a) =1-4 |U_{\a i}|^2(1-|U_{\a i}|^2) 
            \sin^2({\Delta m^2_{31} L \o 4 E}) \ ,
		\end{equation}
 where i=1 or 3 corresponds to hierarchy I or II.
 The mixing parameters can be expressed
 in terms of the oscillation probabilities as \cite{Bil}
\begin{equation}        
 |U_{\a i}|^2={1\o 2}(1\pm \sqrt{1-B_{\n_\a\n_\a}}) \  , 
 \label{dis}
\end{equation}
\noindent with
\begin{equation}        
B_{\n_\a\n_\a}=\{1-P(\n_\a\A \n_\a)\}\sin^{-2}({\Delta m^2_{31} L \o 4 E})\ ,  
 \label{with}
\end{equation}
\noindent where $\a=e$ or $\m$ and  $i=1$ or $3$.
Therefore the parameters $U_{\a i}^2$ at the fixed value of 
$\Delta m^2_{31}$ should satisfy one of the following inequalities: 
\begin{equation}        
 |U_{\a i}|^2 \geq {1\o 2}(1 + \sqrt{1-B_{\n_\a\n_\a}})\equiv a_\a^{(+)} \ ,
    \qquad {\rm or} \qquad
 |U_{\a i}|^2 \leq {1\o 2}(1 - \sqrt{1-B_{\n_\a\n_\a}})\equiv a_\a^{(-)} \ . 
 \label{cons}
 \end{equation}
\noindent The negative results of  Bugey \cite{Bugey}, CDHS \cite{CDHS} 
  and CCFR\cite{CCFR} experiments have given the values of $a_e^{(\pm)}$ and
$a_\m^{(\pm)}$, which were presented in ref. \cite{Bil} and \cite{tani}.
\par
 It is noticed from eq.(\ref{cons}) there are three allowed regions of 
$|U_{e i}|^2$
 and $|U_{\m i}|^2$ as follows:
\begin{eqnarray}  
 (A)\quad |U_{e i}|^2 \geq a_e^{(+)} \ , \qquad  |U_{\m i}|^2 \leq  
 a_\m^{(-)} \ ,
    \nonumber \\
 (B)\quad |U_{e i}|^2 \leq a_e^{(-)} \ , \qquad |U_{\m i}|^2 \leq  
 a_\m^{(-)} \ ,
      \label{ABC} \\     
 (C) \quad |U_{e i}|^2 \leq a_e^{(-)} \ , \qquad |U_{\m i}|^2 \geq  a_\m^{(+)}
 \ ,
    \nonumber
\end{eqnarray}
\noindent
 where $i=1$(hierarchy I) or $3$(hierarchy II).
 In addition to these constraints,  we should take account of
 the constraints by  E531 \cite{E531} and E776 \cite{E776} experimental data.
 These constraints often become  severer than
  the ones of the disappearance experiments as discussed in the next section.
\par
 It may be important to comment on the case (A) with hierarchy I.
  In this case, one has $U_{e3}\simeq 1$ and then the survival probability
 of the solar neutrinos is too large to be consistent with the data of
 GALLEX and SAGE, which have shown less neutrino deficit than the Homestake
  and Kamiokande experiments \cite{solar}.
 Therefore, this case is an unrealistic one for  the neutrino mixings
 although we include this case in our analyses.
\vskip 0.2 cm

 \section{Upper Bound of $J_{CP}^{\n}$} 
 
  The $CP$ violating measure $J_{CP}^{\n}$ defined in eq.(\ref{JCP})
   is also expressed as
 \begin{equation}
   J_{CP}^{\n}= |U_{e 1}||U_{e 2}||U_{e 3} ||U_{\m 3}||U_{\tau 3}|
      (|U_{\m 3}|^2+|U_{\tau 3}|^2)^{-1} \sin\phi \ .
	 \end{equation}
   \noindent
    This formula is rather suitable for  hierarchy II since 
	the experimental constraints are directly  given for $|U_{e 3}|$ 
 and $|U_{\m 3}|$ 
	 as seen in eq.(\ref{ABC}).
   For  hierarchy I, the constraints for $|U_{e 3}|$ and $|U_{\m 3}|$ 
    are indirectly given by using unitarity of the mixing matrix. 
   We discuss the upper bound of $J_{CP}^{\n}$ 
   in six cases: cases (A), (B) and (C) with  hierarchy I or II.
   At first, we study the cases with hierarchy II 
   since those are easier for us 
       to estimate $J_{CP}^{\n}$ than the cases with hierarchy I.\par
   
   The mixing matrix  with  hierarchy II is  written for case (A) as
\begin{equation}  
     {\bf U} \simeq \left (\matrix{ \e_1 &\e_2 &1 \cr 
            U_{\m 1} &U_{\m 2}& \e_3\cr
            U_{\tau 1} &U_{\tau 2}& \e_4 \cr} \right ) \ ,\quad
\end{equation}
\noindent where  $\e_i(i=1\sim 4)$ are tiny  numbers.
 Then  $J_{CP}^{\n}$ is given by 
 \begin{equation}
  J_{CP}^{\n}=\e_1 \e_2 \e_3 \e_4 (\e_3^2 +\e_4^2 )^{-1} \sin\phi
    \leq {1\o 2}\e_1 \e_2 \ ,
 \end{equation}
 \noindent where
  the sign of equality is obtained at $\e_3=\e_4$ with $\sin\phi=1$. 
  Here $\e_3$ is bounded by E776  $\n_\m\A \n_e$ experiment \cite{E776} and
  $\e_4$ is given by unitarity.
   The product  $\e_1 \e_2$ is 
     bounded by unitarity such as 
	
	\begin{equation}
    \e_1 \e_2 \leq {1\o 2}(\e_1^2+ \e_2^2)={1\o 2}(1-|U_{e 3}|^2)
	   \leq {1\o 2} (1-a_e^{(+)}) \ .
   \end{equation} 
 \noindent Thus the upper bound of $J_{CP}^{\n}$ is given only by $a_e^{(+)}$.
In this case   the atomospheric neutrino anomaly could be  attributed to the
   $\n_\m\A \n_\tau$ oscillation 
     if  $|U_{\m 1}|=|U_{\m 2}|=|U_{\tau 1}| =|U_{\tau 2}|\simeq 1/\sqrt{2}$.
	 But, it is emphasized that the estimated upper bound of
   $J_{CP}^{\n}$ is  independent of this condition.\par
	 
	For the case (B)  with  hierarchy II
	 the mixing matrix is given as
   
   \begin{equation}  
  {\bf U} \simeq \left (\matrix{ U_{e 1} &U_{e 2} &\e_1\cr 
            U_{\m 1} &U_{\m 2} & \e_2 \cr
             \e_3 & \e_4 & 1 \cr} \right ) \ .
 \end{equation}
 \noindent
   We get the bound of $J_{CP}^{\n}$ as follows:
 \begin{equation}
  J_{CP}^{\n}= |U_{e 1}||U_{e 2}|\e_1 \e_2  \sin\phi
    \leq {1\o 2}\e_1 \e_2 \ ,
 \end{equation}
 \noindent where
  the sign of equality is obtained at $|U_{e 1}|=|U_{e 2}|= 1/\sqrt{2}$
   with $\sin\phi=1$.  Then the atomospheric neutrino anomaly could be
  solved by the large $\n_\m\A \n_e$ oscillation.
 The bound of $\e_1$  is given by $a_e^{(-)}$ in eq.(\ref{ABC}).  
 On the other hand,
      $\e_2$  is bounded  by  E531  $\n_\m\A \n_\tau$ experiment \cite{E531}
	 since the relevant  transition	 probabilities in the short baseline 
 experiments are given for  hierarchy II:
 \begin{eqnarray}       
 P(\n_\m\A \n_e) \simeq  4 |U_{e 3}|^2 |U_{\m 3}|^2 
            \sin^2({\Delta m^2_{31} L \o 4 E}) \ ,  \nonumber \\
  P(\n_\m\A \n_\tau) \simeq 4 |U_{\m 3}|^2 |U_{\tau 3}|^2 
            \sin^2({\Delta m^2_{31} L \o 4 E}) \ .
	\label{short}
\end{eqnarray}   
  It may be useful to comment on the possibility of 
  the atomospheric neutrino anomaly 
   by the large $\n_\m\A \n_e$ oscillation. 
   The reactor experiments at Bugey \cite{Bugey}  
  and Krasnoyarsk \cite{Krasnoyarsk} have already excluded some large
    $\n_\m - \n_e$ mixing region.
   The allowed one  is 
    $\sin^2 2\th_{e\m}\leq 0.7$ in the case of $\Delta m^2_{21}=10^{-2}\eV^2$.
	 On the other hands, the data of the atomospheric neutrino anomaly
 in Kamiokande \cite{Kam} suggests $\Delta m^2_{21}=7\times 10^{-3}
 \sim 8\times 10^{-2} \eV^2$  
  and $\sin^2 2\th_{e\m}= 0.6\sim 1$ for the  $\n_\m\A \n_e$ oscillation.
  The  overlap region is rather small such as 
  $\sin^2 2\th_{e\m}= 0.6\sim 0.7$. Since 
  the first long baseline reactor experiment CHOOZ \cite{CHOOZ} will  soon give
  the  severer constraint for the $\n_\m -\n_e$ mixing,
   one can check the possibility of the atomospheric neutrino anomaly 
   due to the large $\n_\m\A \n_e$ oscillation.\par
  
  In the case (C)  with  hierarchy II
	 the mixing matrix is 
   
 \begin{equation}  
 {\bf U} \simeq \left (\matrix{ U_{e 1} & U_{e 2} & \e_1 \cr 
            \e_2 & \e_3 & 1 \cr
             U_{\tau 1} &U_{\tau 2}&\e_4 \cr} \right ) \ .
\end{equation}
 \noindent
   Then  we have 
 \begin{equation}
  J_{CP}^{\n}= |U_{e 1}||U_{e 2}|\e_1 \e_4  \sin\phi
    \leq {1\o 2}\e_1 \e_4 \ ,
 \end{equation}
 \noindent where
  the sign of equality is obtained at $|U_{e 1}|=|U_{e 2}|= 1/\sqrt{2}$
   with $\sin\phi=1$.  In this case the atomospheric neutrino anomaly cannot be
  solved by the large $\n_\m$ oscillation because both $\e_2$ and  
 $\e_3$ are very small.
 Here $\e_1$  is bounded by E776  $\n_\m\A \n_e$ experiment \cite{E776}
  while $\e_4$  is  by  E531  $\n_\m\A \n_\tau$ experiment \cite{E531}
   as seen in eq.(\ref{short}).\par
   
   Let us study the cases with hierarchy I, in which
     the relevant  transition  probabilities in the short baseline experiments 
	are given instead of eq.(\ref{short}) as follows:

\begin{eqnarray}   
     P(\n_\m\A \n_e) \simeq  4 |U_{e 1}|^2 |U_{\m 1}|^2 
            \sin^2({\Delta m^2_{31} L \o 4 E}) \ ,  \nonumber \\
     P(\n_\m\A \n_\tau) \simeq 4 |U_{\m 1}|^2 |U_{\tau 1}|^2 
            \sin^2({\Delta m^2_{31} L \o 4 E}) \ .
\end{eqnarray} 
For the case (A) with  hierarchy I
	 the mixing matrix is 
\begin{equation}  
 {\bf U} \simeq \left (\matrix{ 1 &\e_1 &\e_2 \cr 
            \e_3 &U_{\m 2}&U_{\m 3}\cr
             \e_4 &U_{\tau 2}&U_{\tau 3} \cr} \right ) \ .
\end{equation}
\noindent
Then  $J_{CP}^{\n}$ is given by 
 \begin{equation}
  J_{CP}^{\n}=\e_1 \e_2 |U_{\m 3}||U_{\tau 3}|
   (|U_{\m 3}|^2+|U_{\tau 3}|^2)^{-1} \sin\phi
    \leq {1\o 2}\e_1 \e_2 \ ,
 \end{equation}
 \noindent where
  the sign of equality is obtained at $|U_{\m 3}|=|U_{\tau 3}|= 1/\sqrt{2}$
   with $\sin\phi=1$.
   The product  $\e_1 \e_2$ is bounded by unitarity such as 
	
	\begin{equation}
    \e_1 \e_2 \leq {1\o 2}(\e_1^2+ \e_2^2)={1\o 2}(1-|U_{e 1}|^2)
	   \leq {1\o 2} (1-a_e^{(+)}) \ .
   \end{equation} 
\noindent Thus we get  the same  upper bound of $J_{CP}^{\n}$ as the one in the
    case (A) with hierarchy II.
	The atomospheric neutrino anomaly could be  attributed to the
   $\n_\m\A \n_\tau$ oscillation 
  if  $|U_{\m 2}|=|U_{\m 3}|=|U_{\tau 2}| =|U_{\tau 3}|\simeq 1/\sqrt{2}$.\par
 
For the case (B)  with  hierarchy I
	 the mixing matrix is written as
\begin{equation}  
 {\bf U} \simeq \left (\matrix{ \e_1 &U_{e 2}&U_{e 3}\cr 
            \e_2 &U_{\m 2}&U_{\m 3}\cr
             1 & \e_3 & \e_4 \cr} \right ) \ .
\end{equation}
\noindent
Then  we get
 \begin{equation}
  J_{CP}^{\n}=\e_1 \e_4 |U_{e 2}||U_{e 3}||U_{\m 3}|
   (|U_{\m 3}|^2+\e_4^2)^{-1} \sin\phi
    \leq {1 \o \sqrt{2}}\e_1 \e_4 \ ,
	\label{bound}
 \end{equation}
 \noindent where
  the sign of equality is obtained at 
  $|U_{e 2}|=|U_{e 3}|=|U_{\m 3}|= 1/\sqrt{2}$
   with $\sin\phi=1$.
   However, this bound is not exact one in contrast to previous cases.
 We checked  numerically that eq.(\ref{bound}) gives roughly the  maximum value
    by using the present bound of  $\e_4$.
  The magnitude of $\e_1$  is bounded  by $a_e^{(-)}$ in eq.(\ref{ABC}) while
      $\e_4$ is bounded by unitarity such as 
 
 \begin{equation}
   \e_4^2 = \e_1^2+\e_2^2-\e_3^2 \leq \e_1^2+\e_2^2  \ ,
   \end{equation} 
   \noindent where $\e_2$ is bounded
     by  E531  $\n_\m\A \n_\tau$ experiment \cite{E531}.
	  The   upper bound of $J_{CP}^{\n}$ is different from the one in
    the case (B) with hierarchy II.
	The atomospheric neutrino anomaly could be
  solved by the large $\n_\m\A \n_e$ oscillation.

For the case (C)  with  hierarchy I
	 the mixing matrix is 
\begin{equation}  
 {\bf U} \simeq \left (\matrix{ \e_1 & U_{e 2} &U_{e 3} \cr 
            1 & \e_2 & \e_3 \cr
             \e_4 &U_{\tau 2}&U_{\tau 3} \cr} \right ) \ .
\end{equation}
\noindent   
Then  $J_{CP}^{\n}$ is given by 
 \begin{equation}
  J_{CP}^{\n}=\e_1 \e_3 |U_{e 2}||U_{e 3}||U_{\tau 3}|
   (\e_3^2+|U_{\tau 3}|^2)^{-1} \sin\phi
    \leq {1\o \sqrt{2}}\e_1 \e_3 \ ,
 \end{equation}
 \noindent where
  the sign of equality is obtained at 
  $|U_{e 2}|=|U_{e 3}|=|U_{\tau 3}|\simeq 1/\sqrt{2}$
   with $\sin\phi=1$.
    This bound is  also not exact one as well as the case (B)
	 although it is  roughly the  maximum value.
 Here $\e_1$  is bounded by E776  $\n_\m\A \n_e$ experiment \cite{E776}.
  On the other hand, $\e_3$ is bounded by unitarity

 \begin{equation}
  \e_3^2= |U_{\m 3}|^2 = 1-|U_{\m 1}|^2-|U_{\m 2}|^2 \leq 1-a_{\m}^{(+)} \ ,
   \end{equation} 
   \noindent where
   $a_{\m}^{(+)}$ is given by the disappearance experiments as seen in eq.(\ref{ABC}). 
 The   upper bound of $J_{CP}^{\n}$ is different from the one in
    the case (C) with hierarchy II. \par

Thus we obtain the upper bounds of $J_{CP}^{\n}$ for six cases 
 which are allowed
 by the present short baseline experiments.
 \vskip 0.5 cm
 \section{Numerical Results of $CP$ Violation} 
 
 Now we can calculate the upper bound of 
 $\Delta P\equiv P(\n_\m \A \n_\tau) - P(\bar \n_\m \A \bar \n_\tau)$, which is
 the direct measurement of  $CP$ violation.
  Since the upper bounds of $J_{CP}^{\n}$ have been given  
      for fixed $\Delta m_{31}^2$,
  the   upper bounds of $\Delta P$ are also presented  for $\Delta m_{31}^2$ 
   with fixing $L$, $E$ and  $\Delta m_{32}^2$( for hierarchy I) or 
   $\Delta m_{21}^2$( for hierarchy II).
   In fig.2(a), we show numerical results for cases (A), (B) and (C) 
     of  hierarchy I 
   with  $L=250$Km and  $\Delta m_{32}^2= 10^{-2}\eV^2$.
  Fig.2(b) corresponds to hierarchy II.
  Here we used the energy band of 
 $E=1 \sim 1.5\G$ in the energy spectrum of the incident neutrino,
  which is expected in KEK-PS \cite{KEKPS}.
 Then we get the averaged value $S_{CP}=0.725$ for 
  $\Delta m_{31}^2= 1\sim 25 \eV^2$, which is used in our
 calculation to avoid long CPU time due to the oscillatory integrand.
 Therefore, one should take into consider $5\%$ error
 in the results of Figs. (a) and (b).
  
\begin{center}
\unitlength=0.7 cm
\begin{picture}(2.5,1.5)
\thicklines
\put(-2,0){\framebox(6,1){\bf Fig. 2(a) and (b)}}
\end{picture}
\end{center}
\par
The weakest bound is given in the case (B) with hierarchy I,
in which the bound is almost determined only by Bugey
 reactor disappearance experiment \cite{Bugey}.
 In this case $\Delta P$ could be  $10^{-1}$,
  which can be observed in KEK-SuperKamiokande experiment.
  Then the atomospheric neutrino anomaly is due to
	  the large $\n_\m\A \n_e$ oscillation.
  The first long baseline reactor experiment CHOOZ \cite{CHOOZ} will
     soon test this possibility	  by presenting the  severer constraint of
  the $\n_\m -\n_e$ mixing.\par
	  
	 The observation of the $CP$ violating effect is not expected
	  in the case (C) for both hierarchies I and II
        since the upper bounds are
	   around or below $10^{-2}$. 
	    In addition, the atomospheric neutrino anomaly could not be
	   explained by the large neutrino mixing.\par
	   
	If the large $\n_\m\A \n_\tau$ oscillation causes 
	the atomospheric neutrino anomaly, the case (A) is prefered.
	The upper bound is  around $0.03$, which is same for both hierarchies.
	Since this  bound is  determined only by the
    reactor disappearance experiments \cite{Bugey},
	it will be improved by new disappearance experiments.\par
	
	It is remarked that the estimated  upper bounds of  $J_{CP}^{\n}$  
 are given by the maximal  mixing such as $|U_{\a i}|\simeq 1/\sqrt{2}$
	  except for the case (A) with hierarchy II.
  If the atomospheric neutrino anomaly is not due to the large neutrino mixing,
	  the $CP$ violating effect is reduced. 
   The situation is  different  in the case (A) with hierarchy II.
	In this case, the upper bound  has been obtained 
	  without assuming the large neutrino mixing.\par
	  
 In our analyses, we do not take account of the new experimental data given 
  by LSND \cite{LSND}.
 Even if the data is included, our obtained bounds do not almost change.

\vskip 0.5 cm

\noindent
 \section{Matter Effect}\par
The general discussion of the matter effect in the long baseline
 experiments  was given by Kuo and Pantaleone \cite{matter}.
 The data in those experiments  include the background
 matter effect  which is not $CP$ invariant.  
 Therefore, it is very important to investigate the matter effect
  in order to estimate the $CP$ violation effect
 originated by the phase of the
 neutrino mixing  matrix. The  matter effect of the earth
 should be carefully analized since the effect considerably
  depends on the mass hierarchy and mixings as well as
 the incident energy of the neutrino.\par

We estimate the matter effect on the transition probabilities
  by   switching off $CP$ violation due to the mixing matrix.
Then,  $\Delta P$ is given by the only matter effect.
 If the estimated $\Delta P$ is comparable to the ones in the
 previous section, one should consider how to
  extract the matter effect from the data.\par

 The matter effect in the long baseline accelerater experiments
       is rather easily estimated by assuming the constant electron density.
The effective mass squared in matter $M_m^2$ for
 neutrino energy $E$ in weak basis \cite{matter} is

\begin{equation}  
 {\bf M_m^2} = U_m \left (\matrix{ m_1^2 & 0 & 0 \cr 
            0 & m_2^2 & 0 \cr
            0 & 0 & m_3^2 \cr} \right )U_m^\dagger +
      \left (\matrix{ A & 0 & 0 \cr 
            0 & 0 & 0 \cr
            0 & 0 & 0 \cr} \right )  \ ,
\end{equation}
\noindent   
where $A=2\sqrt{2} G_F n_e E$. We use the constant
 electron density $n_e=2.4\ {\rm  mol/cm}^3$.
 For antineutrinos, the effective mass squared is given by
 replacing $A\A -A$ and $U_m \A U_m^*$
 The effective mixing matrix without the $CP$ violating phase
 $U_m$ is written by
\begin{equation}  
 {U_m} =  \left (\matrix{ c_{m13} c_{m12} & c_{m13} s_{m12} &  s_{m13} \cr 
  -c_{23}s_{m12}-s_{23}s_{m13}c_{m12} & c_{23}c_{m12}-s_{23}s_{m13}s_{m12} & 
                       s_{23}c_{m13} \cr
  s_{23}s_{m12}-c_{23}s_{m13}c_{m12} & -s_{23}c_{m12}-c_{23}s_{m13}s_{m12} & 
                       c_{23}c_{m13} \cr} \right ) \ ,
\end{equation}       
\noindent
 where $s_{mij}\equiv \sin{\theta^m_{ij}},\ c_{mij}\equiv \cos{\theta^m_{ij}}$
 for  effective mixings in the matter and
  $s_{ij}\equiv \sin{\theta_{ij}},\ c_{ij}\equiv \cos{\theta_{ij}}$
 for vacuum mixings.
 For example,  the effective mixing angle $s_{m12}$ is given 
 in terms of vacuum mixings as
\begin{equation}  
  \sin 2 \theta^m_{12} = {\Delta m^2_{21}\sin 2\theta_{12}  \o
  \sqrt{(A \cos^2\theta_{13}- \Delta m^2_{21}\cos 2\theta_{12})^2+
  \Delta m^2_{21}\sin^2 2\theta_{12} }} 
\end{equation}       
\noindent to zeroth order in $A \sin 2\theta_{13}$.

In the case of hierarchy I($m_3\simeq m_2 \geq 1\eV$),
 the matter effect is expected to be small
 because $ m^2_{21} \gg A\simeq 5 \times 10^{-4}\eV^2$.
  In fact, the obtained  $\Delta P$ is at most $5\times 10^{-3}$
  for three cases of (A), (B) and (C).
Therefore,  the matter effect do not disturb  the
 information of  $CP$ violation originated by
 the $CP$ violating phase in the neutrino mixing matrix
       if  the observed $\Delta P$
 is not far from our estimated upper bounds in section 6(see Fig.2).

  However, the case with  hierarchy I is another.
Since  the value of $A$ is not negligible compared to
 $m^2_{21} \simeq  10^{-2}\eV^2$, the matter effect is expected to 
 be important.
 We show the matter effect $\Delta P$ versus $s_{12}$
 for the typical parameters    in Figs.3(a) and (b).
  Here
  the solid curves denote the matter effect  $\Delta P(\n_\m\A \n_\tau)$
 and the dashed curves denote  $\Delta P(\n_\m\A \n_e)$,
  where $\Delta P(\n_\a\A \n_\b)\equiv 
   P(\n_\a\A \n_\b) -  P(\bar \n_\a\A \bar \n_\b)$
                 \footnote{
 The $CP$ non-invariant quantity   
 $|\Delta P(\n_\m\A \n_\tau)|$ is different from $|\Delta P(\n_\m\A \n_e)|$
  for the matter effect.  On the other hand,
  those have  same magnitudes for  $CP$ violation originated by the
 phase in the neutrino mixing  matrix as seen in Eq.(2).}.
 The parameters are fixed to be 
 $m^2_{31}=2.25\eV^2$,  $m^2_{21}=10^{-2}\eV^2$,  $L=250 {\rm Km}$
 and  $E=1.2\G$.
 The vacuum mixing angles are taken to be
  $s_{13}=0.96$ and $s_{23}=1/\sqrt{2}$ for case (A)
 and  $s_{13}=0.15$ and $s_{23}=0.12$ for case (B).

\begin{center}
\unitlength=0.7 cm
\begin{picture}(2.5,1.5)
\thicklines
\put(-2,0){\framebox(6,1){\bf Fig. 3(a) and (b)}}
\end{picture}
\end{center}

 In case (A),  the $\n_\m - \n_\tau$ mixing is maximum
 at $s_{12}=0$ or $1$. Then, the matter effect  $\Delta P(\n_\m\A \n_\tau)$
  increases up to $7.5 \times 10^{-3}$. For the $\n_\m - \n_e$
 mixing which  is very small in this case, the matter effect 
  $\Delta P(\n_\m\A \n_e)$ is tiny.
  If  the  $CP$ violation effect $\Delta P$
  is larger than  $10^{-2}$ as shown in Fig. 2(b), 
  the matter effect does not dominate  $\Delta P$.
 The smallness of the matter effect is due to
 the suppressed $A c_{13}^2$ in eq.(31)
 ($c_{13} \ll 1$ in case (A)). \par

In case (B), the $\n_\m - \n_e$ mixing is maximum at
 $s_{12}=1/\sqrt{2}$. The matter effect  $\Delta P(\n_\m\A \n_e)$
  could be  $8 \times 10^{-2}$. Since the $\n_\m - \n_\tau$
 mixing is very small in this case, the matter effect 
  $\Delta P(\n_\m\A \n_\tau)$ is also small.
 Since the matter effect is very large compared to the results
 in Fig.2(b), it is difficult to get the information of the
 $CP$ violating phase  in the mixing matrix. 
\par

 In case (C), both $\n_\m - \n_e$ and $\n_\m - \n_\tau$
 mixings are very small, and so the matter effects $\Delta P$'s are also
 small, at most $5\times 10^{-4}$. In this case, there is no
 hope to observe the neutrino oscillation in the planned
 long baseline   accelerator  experiments.\par

Thus, the matter effect becomes important to observe
 $CP$ violation in the case of hierarchy II.
 However, recent works of the pattern of the neutrino masses and
 mixings \cite{sterile}\cite{pattern} may exclude hierarchy II.
If hierarchy I is realized for the  neutrino masses, 
  the matter effect does not almost modify
 our results in Fig.2(a).

\noindent
 \section{Conclusions}\par

 We have studied the direct measurements of  $CP$ violation
 originated by the phase of the neutrino mixing matrix
 in the long baseline  neutrino oscillations.
	In those experiments, the $CP$ violating effect is not suppressed 
	 if the highest neutrino mass scale is 
	  taken to be $1\sim 5\eV$, which is appropriate for the cosmological
  HDM.
 The upper bounds have been calculated  for three cases (A), (B), (C)
 in hierarchies I and II, where  mixings
	 are constrained by the recent short baseline ones.  
   The estimated upper bounds are
    larger than  $10^{-2}$, which is observable
	 in the long baseline accelerator experiments.
 The new reactor disappearance experiments will provide severer
	 bound in the near future.

 The matter effect on $CP$ violation is also calculated.
  The effect is not significant for hierarchy I, but for hierarchy II.
  The recent works of the neutrino masses and mixings suggest
 the case (A) with hierarchy I.
 In this case, the matter effect on  $CP$ violation is negligible
  if the observed $\Delta P$ is close to our estimated
 upper bound.

\vskip 3 cm
\centerline{ \Large \bf Acknowledgments}
I would like to thank J.  Pantaleon for his critical comments on
   the matter effect. 
 I  thank  A. Smirnov and H. Nunokawa for the  quantitative
 discussion of the matter effect.
 I also  thank  S.M. Bilenky and W. Grimus
  for discussing the $CP$ violating experiments.
 I  wish to acknowledge the hospitality of DESY theory group,
 in particular, A. Ali.
This research is  supported by Alexander von Humboldt foundation 
 (Germany) and
 the Grant-in-Aid for Science Research,
Ministry of Education, Science and Culture, Japan(No. 07640413).
\newpage

\newpage
\topskip 2 cm
\centerline{\Large{\bf Figure Captions}}
\par
\vskip 1.5 cm
\noindent
{\bf Figure 1:}\par
  Dependences of $S_{CP}$ on the  neutrino energy $E$ 
	  for  (a) $L=250$Km and  (b) $L=730$Km
	   with $\Delta m^2_{31}=2.25\eV^2$ and 
  $\Delta m^2_{32}= 10^{-2}\eV^2$,
	    which are shown by solid curves.
	 The dashed curves denote  $\sin^2(\Delta m^2_{32} L/4 E)$.
 Those are averaged over  energy speread of $20\%$ of
 the  neutrino energy.   

\vskip 1 cm
\noindent
{\bf Figure 2:}\par
 Upper bounds of $\Delta P$ versus $\Delta m^2_{31}$
   for (a) hierarchy I and (b) hierarchy II.
   The solid, dashed and dashed-dotted  curves denote  
   the cases (A), (B) and (C),
   respectively.   We take   $E=1\sim 1.5\G$ and $L=250$Km
 with  $\Delta m^2_{32}= 10^{-2}\eV^2$ for hierarchy I and
  $\Delta m^2_{21}= 10^{-2}\eV^2$ for hierarchy II.
\vskip 1 cm

\noindent
{\bf Figure 3:}\par
  The matter effect  $\Delta P$ versus $s_{12}$
   in  hierarchy II for (a) case (A) and (b) case (B).
   The solid and  dashed  curves denote  
      $\Delta P(\n_\m\A \n_\tau)$ and   $\Delta P(\n_\m\A \n_e)$,  
   respectively.   Here  $\Delta m^2_{31}= 2.25\eV^2$,
 $\Delta m^2_{21}=10^{-2}\eV^2$,  $E=1.2\G$
   and $L=250$Km are taken.

\newpage
\epsfig{file=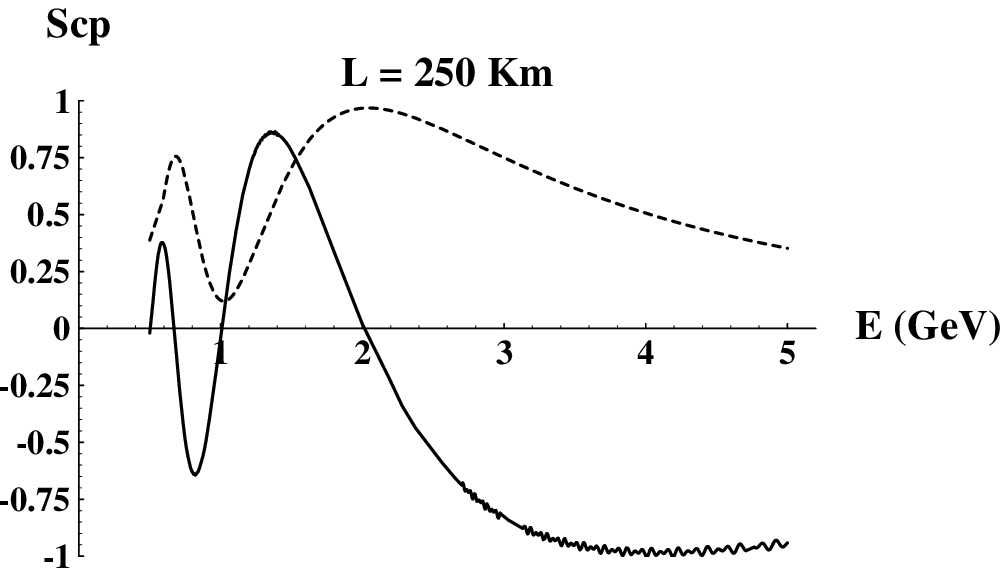, width=14 cm}

  \centerline{\large \bf Fig. 1 (a) }

\epsfig{file=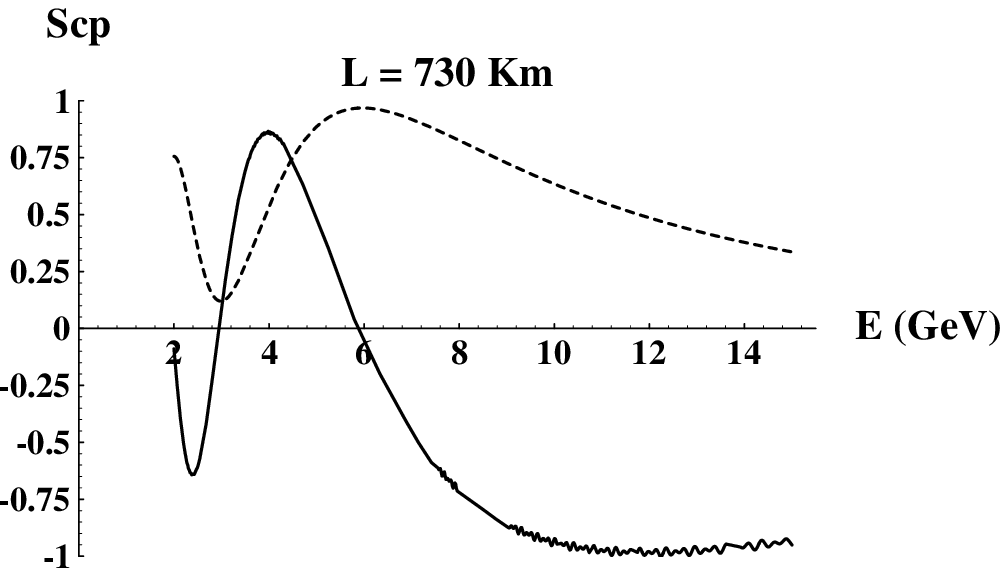, width=14 cm}
 
 \centerline{\large \bf Fig. 1 (b) }

\epsfig{file=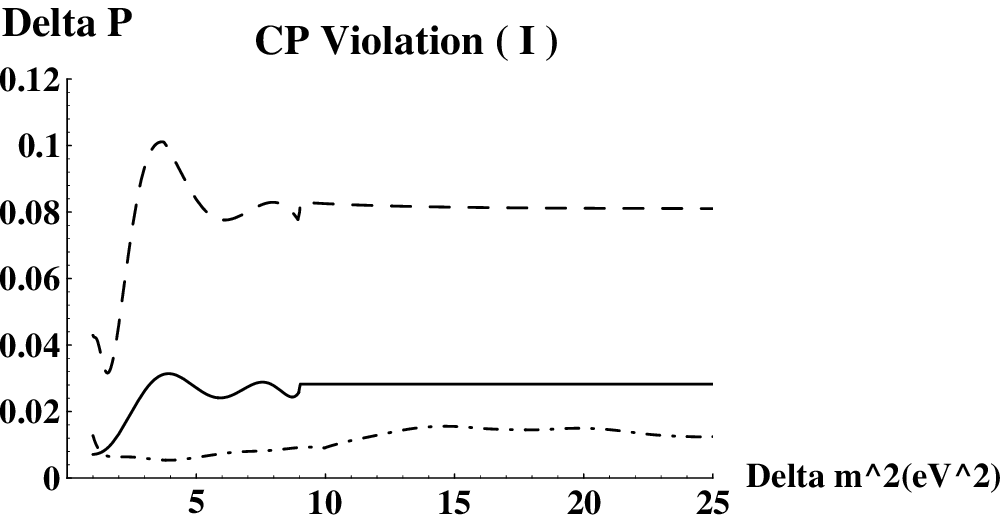, width=14 cm}

  \centerline{\large \bf Fig. 2 (a) }

\epsfig{file=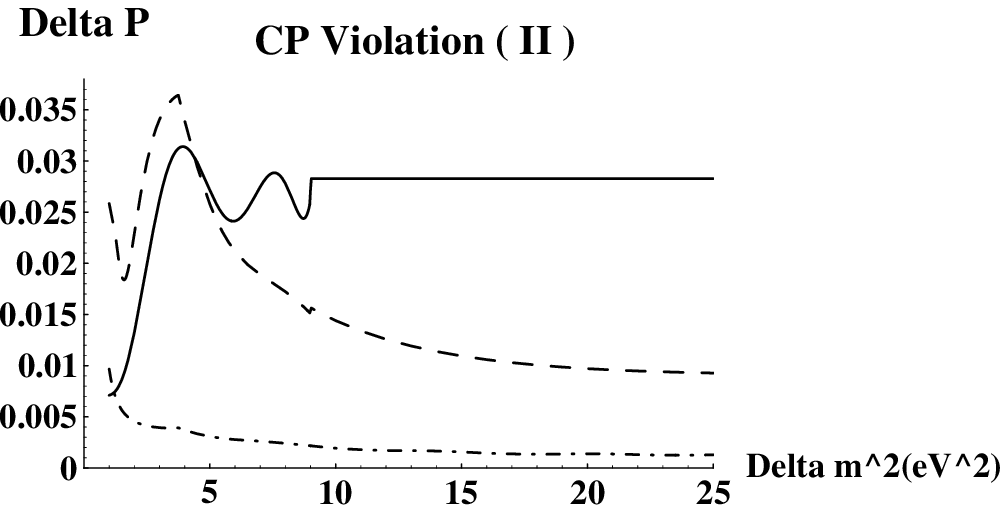, width=14 cm}
 
 \centerline{\large \bf Fig. 2 (b) }

\epsfig{file=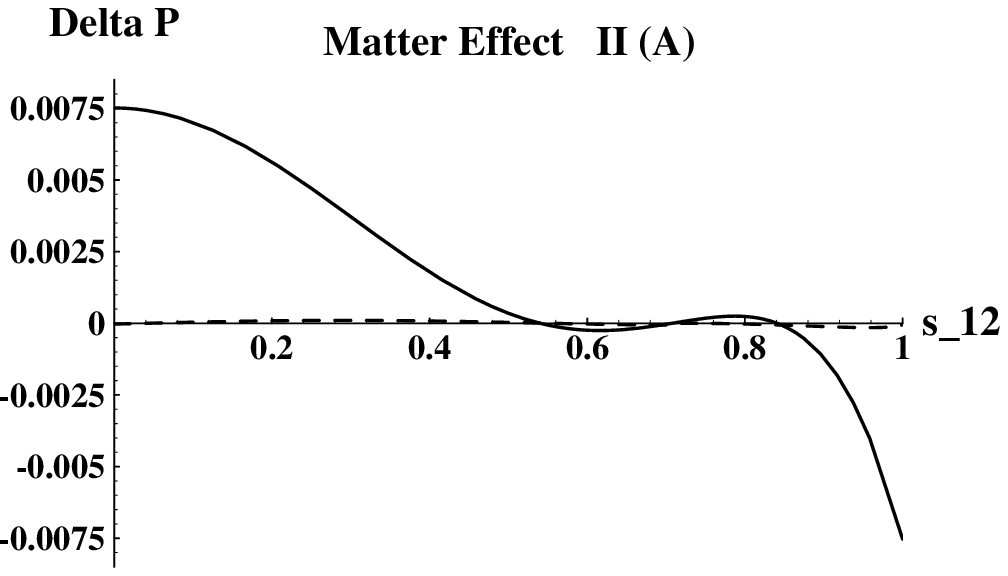, width=14 cm}

  \centerline{\large \bf Fig. 3 (a) }

\epsfig{file=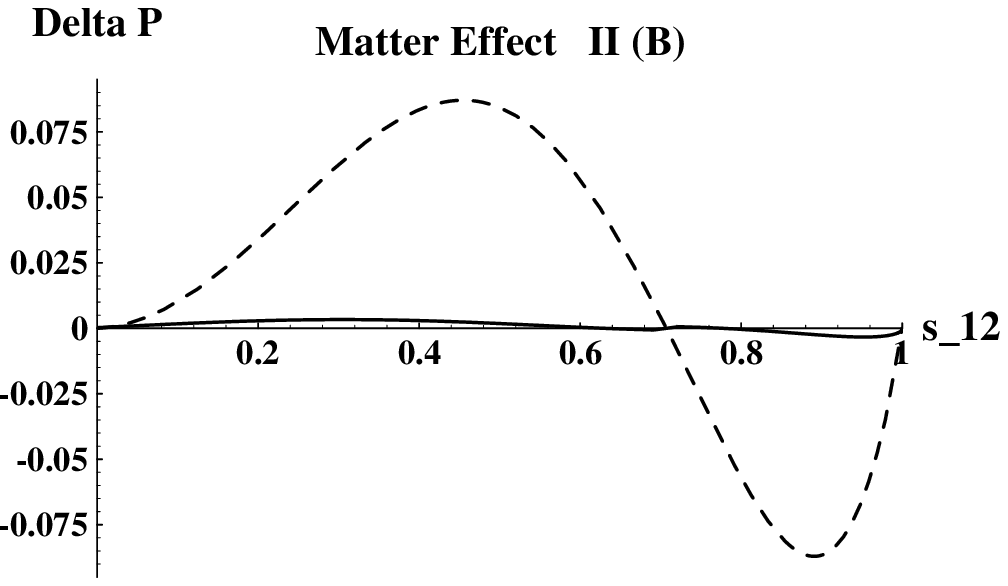, width=14 cm}
 
 \centerline{\large \bf Fig. 3 (b) }

\end{document}